\documentclass{timproc}

\usepackage[utf8]{inputenc}

\usepackage[figuresright]{rotating}

\usepackage{hyperref}

\usepackage{graphicx}	
\usepackage{amsmath}	
\usepackage{amssymb}	
\usepackage{slashbox}
\usepackage{color}
\usepackage{mathptmx} 
\usepackage{multirow}
\usepackage{wasysym}
\usepackage{url}
\usepackage{natbib}

\def\apjl{ApJ}
\def\apjs{ApJS}
\def\ao{Appl.~Opt.}
\def\aap{A\&A}

\title[Water transport towards circumprimary HZ]{Water transport to circumprimary habitable zones from icy planetesimal disks in binary star systems}

\author[Bancelin, Pilat-Lohinger, Maindl \& Bazs\'o]{D. Bancelin, E. Pilat-Lohinger, T.~I. Maindl and \'A. Bazs\'o}
\affiliation{University of Vienna, Department of Astrophysics, Türkenschanzstr. 17,\\ 1180 Vienna, Austria\\
	\email{david.bancelin@univie.ac.at}
}

\begin{document}
\setcounter{page}{21}
\maketitle

\begin{abstract}
So far, more than 130 extrasolar planets have been found in multiple stellar systems. Dynamical simulations show that the outcome of the planetary formation process can lead to different planetary architectures (i.e. location, size, mass, and water content) when the star system is single or double. In the late phase of planetary formation, when embryo-sized objects dominate the inner region of the system, asteroids are also present and can provide additional material for objects inside the habitable zone (HZ). In this study, we make a comparison of several binary star systems and aim to show how efficient they are at moving icy asteroids from beyond the snow line into orbits crossing the HZ. We also analyze the influence of secular and mean motion resonances on the water transport towards the HZ. Our study shows that small bodies also participate in bearing a non-negligible amount of water to the HZ. The proximity of a companion moving on an eccentric orbit increases the flux of asteroids to the HZ, which could result in a more efficient water transport on a short timescale, causing a heavy bombardment. In contrast to asteroids moving under the gravitational perturbations of one G-type star and a gas giant, we show that the presence of a companion star not only favors a faster depletion of our disk of planetesimals, but can also bring 4–5 times more water into the whole HZ. However, due to the secular resonance located either inside the HZ or inside the asteroid belt, impacts between icy planetesimals from the disk and big objects in the HZ can occur at high impact speed. Therefore, real collision modeling using a GPU 3D-SPH code show that in reality, the water content of the projectile is greatly reduced and therefore, also the water transported to planets or embryos initially inside the HZ.

\end{abstract}

\section{Introduction}
Nearly 130 extra solar planets in double and multiple star systems have been discovered to date. Roughly one quarter of these planets are orbiting close to or even crossing their systems' habitable zone (HZ), i.e. the region where an Earth-analogue
could retain liquid water on its surface \citep{rein14}. While most of these planets are gas giants, the incredible ratio of one in four planets being at least partly in the HZ seems to make binary star systems promising targets for the search of a second Earth, especially for the next generation of photometry missions CHEOPS, TESS, and PLATO-2.0. About 80 percent \citep{rein14} of the currently known planets in double star systems are in so-called S-type configurations \citep{dvorak84},  i.e. the two stars are so far apart so that the planet orbits only one stellar component without being destabilized. As many of the wide binary systems host more than one gas giant, their dynamical evolution is quite complex. The question whether habitable worlds 
can actually exist in such environments is, therefore, not a trivial one. Previous works on early stages of planetary formation have shown that planetesimal accretion can be more difficult than in single star systems \citep[and references therein]{thebault14}. This in turn can question the possibility of whether  embryos form in such systems. However, studies of late stages of planetary formation show that, should embryos manage to form despite these adverse conditions,  the dynamical influence of companion stars is not prohibitive to forming Earth-like planets \citep{raymond04,haghighipour07}. Furthermore, it was shown that binary star systems in the vicinity of the solar system are capable of sustaining habitable worlds once they are formed \citep{eggl13,jaime14}. As the amount of water on a planet's surface seems to be crucial to sustaining a temperate environment \citep{kasting93}, it is important to identify possible sources. For Earth, two mechanisms seem to be important: i) endogenous outgassing of primitive material and ii) exogenous impact
by asteroids and comets sources. Since neither can explain the amount of and isotope composition of Earth's oceans in itself, models that favour a combination 
of both sources seem to be more successful \citep{izidoro13}. The amount of primordial water that is collected during formation phases of planets in S-type orbits in binary 
star systems containing additional gas giants has been studied by \cite{haghighipour07}. They have shown that the planets formed in a circumstellar HZ may have collected between 4 and 40 Earth oceans from planetary embryos, but a main trend  appeared: the more eccentric the orbit of the binary is, the more eccentricity is also injected into the gas giant's orbit. This in turn leads to fewer and dryer terrestrial planets. Stochastic simulations proved that almost dry planets can also be formed  in the circumprimary HZ of binary star systems \citep{haghighipour07}. However, as emphasized by these authors, water delivery in the inner solar system is 
not only due to radial mixing of planetary embryos. Smaller objects can also contribute as shown in \cite{raymond07}.

Indeed, mean motion (MMRs) and secular (SR) resonances play a key role in the architecture of a planetary system. It is well known that they had a strong influence on the dynamics in the early stage \citep{walsh11} and late stage \citep{tsiganis05,gomes05} of the planetary formation in our solar system. Icy bodies trapped into orbital resonances could be potential water sources for planets in the HZ. These water rich objects can be embryos and small bodies (asteroids) as shown in \cite{morbidelli00} and \cite{obrien14} for our Earth.

In this study, we aim to answer how much water can be transported into the HZ via small bodies thus providing additional water sources to objects orbiting in the HZ.
In Sect. \ref{S:statistical}, we study statistically the dynamics of a circumprimary asteroid belt in some binary star configurations \citep{bancelin15}. We treat this problem in a self-consistent manner as all gravitational interactions in the system as well as water loss of the planetesimals due to outgassing are accounted for. Then, in Sect. \ref{S:dynamics}, we aim to emphasize and characterize the dynamical effects of orbital resonances on a disk of planetesimals, in various binary star systems hosting a gas giant planet, as well as to what extent such resonances are likely to enable icy asteroids to bring water material into the HZ in comparison to single star systems \citep{bancelin16}. For various binary star -- giant planet configurations, we investigate in detail in Sect. \ref{S:EP} the influence of the secular resonance, located $\sim$ 1.0 au, on the water transport to bigger objects (embryos or planets) orbiting the host-star within the HZ \citep{bancelin17}. Finally, we conclude our work in Sect. \ref{S:conclusion}. 

\section{Water transport: statistical overview} \label{S:statistical}

\subsection{Initial modelling}

We focus this study on binary star systems with two G-type stars with masses equal to one solar mass. The initial orbital separations are either $a_{\scriptscriptstyle \text{b}}$ = 50 au or 100 au and the binary's eccentricity is $e_{\scriptscriptstyle \text{b}}$ = 0.1 or 0.3 (see also \cite{bancelin15} for more binary star configurations). Our studied systems host a gas giant planet initially at a semi-major axis $a_{\scriptscriptstyle \text{GG}}$ = 5.2 au, moving on a circular orbit with a mass equal to the mass of Jupiter. Since we study the planar case, initial inclinations are set to 0$^{\circ}$.

A disk of planetesimals is modelled as a ring of 10000 asteroids with masses 
similar to main belt objects in the solar system and each asteroid was 
assigned an initial water mass fraction (hereafter $wmf$) of 10\%
\citep{abe00,morbidelli00}. To determine the lower and upper limits for the asteroids' masses, we performed independent preliminary simulations with a 3D smooth particle 
hydrodynamics (SPH) code \citep{sch05,maisch13,schrie16}. The scenarios involve collisions of rocky basaltic objects with one lunar mass at different encounter velocities and 
angles (see Tab. 1 in \cite{bancelin15}). Our results show that in a hit-and-run and merging scenarios (see Fig. 1 in \cite{bancelin15}), all of the ten largest fragments possess masses $\ge 1\,\%\/$ of the total system mass, which is approximately  Ceres' mass and hundreds are above the ``significant'' fragment threshold in the sense of 
\citet{maidvo14b}. The smallest fragment consists of one SPH particle (0.001\% of the total 
mass for 100k SPH particles) which corresponds to $\sim$ 0.1\% of Ceres' mass. As increasing the number of SPH 
particles will 
result in even smaller fragments, this mass is an upper limit for the smallest fragment. As this fragment will 
contain $\sim$ 0.006\% Earth-oceans units\footnote{1 ocean = $1.5 \times
10^{\scriptscriptstyle 21}$ kg of H$_{\scriptscriptstyle 2}$O}, we chose to neglect the water contribution of 
smaller particles. Our minimum and maximum mass are thus defined according to the fragments' mass after one impact. Therefore, members of our ring will have masses
randomly and equally distributed between 0.1\% to one Ceres' mass. The total mass of the ring of 10000 asteroids amounts to 0.5 $M_{\oplus}$. Thus, the quantity of water in terms of Earth-ocean units available in a ring will be 200.

We randomly distributed asteroids inside and beyond the orbit of the gas giant. To avoid strong initial interaction with the gas giant, we assumed that it has gravitationally cleared a path in the disk around its orbit. We defined the width 
of this path as $\pm 3\,R_{\scriptscriptstyle {\text{H},\text{GG}}}$, where $R_{\scriptscriptstyle {\text{H},\text{GG}}}$ is the giant planet's Hill radius. The inner border of the disk is set to the snow-line position \citep{lecar06,martin12,martin13}, the border between icy and rocky planetesimals. In case of primary G-type star, the value of the snow-line is 2.7 au.  The outer border is defined according to the critical semi-major axis $a_{\scriptscriptstyle \text{c}}$ and its uncertainty $\Delta
a_{\scriptscriptstyle \text{c}}$ \citep{holman99,pilat02}. All the 
simulations are purely gravitational since we started our numerical integrations to start after the gas has vanished (therefore, we do not consider gas driven migration and eccentricity dampening). The initial eccentricities and inclinations are randomly chosen below 0.01 and 1$^\circ$ respectively. We limited our study to 10 Myrs integration time and integrated numerically our systems using the \textit{nine} package \citep{eggl10}. The numerical integrator used for 
the computations is based on the  Lie-series (see e.g. \cite{hanslmeier84} and more recently \cite{bancelin12}). As our planetesimals do not interact with each other, the disk was divided into 100 sub-rings which were integrated separately. 

\subsection{Statistics of the disk dynamics}
\begin{figure}[!h]
  \centering{
 \includegraphics[angle=-90,width=0.8\textwidth]{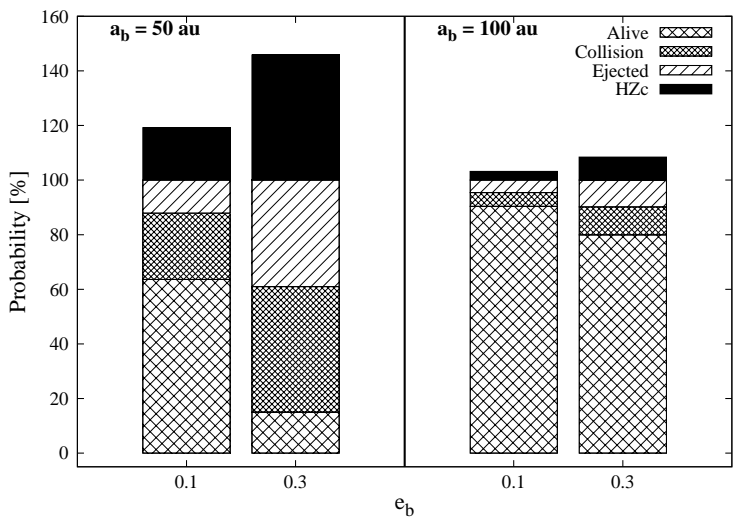}}
  \caption{Statistics on the  dynamics of the disk of planetesimals. Each histogram shows the 
evolution (expressed in probability) of the
asteroids in the belt within 10 Myr of integration. Asteroids can either be still present in the system 
("alive") or become an HZc or collide with
the stars or the gas giant,  or be ejected out of the system. The closer the secondary, the higher is the probability to empty the asteroids ring.}
    \label{F:impactor}
\end{figure}

During the simulation, each particle is tracked until the end of the
integration time in order to assess:
\begin{itemize}

 \item [(a)] asteroids crossing the HZ. They will be referred to as habitable zone crossers (hereafter HZc). As we assume a two-dimensional HZ, an asteroid will be considered as an HZc if the intersection point between its orbit and the HZ plane lies within the HZ borders;
 \item [(b)] asteroids leaving the system when their semi-major axis$\ge$ 500 au;
 \item [(c)] asteroids colliding with the gas giant or the stars and
 \item [(d)] asteroids still alive in the belt after 10 Myr.
\end{itemize}

Figure \ref{F:impactor} shows the resulting statistics on the asteroids' dynamics for $a_{\scriptscriptstyle \text{b}}$ = 50 au (left panel) and 100 au (right panel) (see also Fig. 4 in \cite{bancelin15}). For each semi-major axis, we show the dynamical outcome of our asteroids expressed in terms of probability, as a function of the secondary's eccentricity. Below 100\%, the percentage of asteroids that are still present in the belt (``alive''), that were ejected or that collided with the stars or the gas giant is shown. The black area of each histogram above 100\% indicates the probability that asteroids will enter the HZ i.e. HZc. A comparison of the different histograms indicates that the most important parameter is the periapse of the binary system, which is defined by the semi-major axis and eccentricity of the binary. Figure \ref{F:impactor} clearly shows that the probability of asteroids becoming HZc increases if the secondary's periapsis distance decreases. Indeed, for a given value of 
$a_{\scriptscriptstyle \text{b}}$ (for instance 50 au), one can see that
this probability is at least doubled when $e_{\scriptscriptstyle \text{b}}$ increases. As a consequence, the asteroid belt will be depopulated because of dynamically induced ejections, as well as collisions with the giant planet and the stars. Since the rate of colliding and ejected asteroids is higher, a ring will be depopulated faster when $e_{\scriptscriptstyle \text{b}}$ becomes larger. Therefore, the statistics in Fig. \ref{F:impactor} shows that the probability for an asteroid to remain in the ring after 10 Myrs will decrease with the periapsis distance.

\subsection{Timescale statistics}
Depending on the periapsis distance of the secondary, the disk of planetesimals can be 
perturbed more or less rapidly. Asteroids will suffer from the gravitational perturbations of the 
secondary star and the gas giant, and their eccentricity may increase quickly. 
\begin{figure}[!h]
  \centering{
 \includegraphics[angle=-90, width=0.8\linewidth]{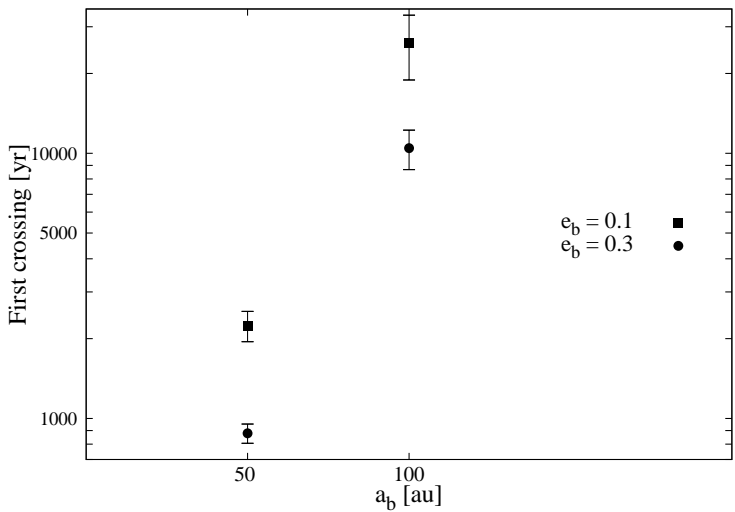}}
  \caption{Median time for an asteroid to become an HZc for the case of a secondary with 
e$_{\scriptscriptstyle \text{b}}$ = 0.1 ($\blacksquare$) and e$_{\scriptscriptstyle \text{b}}$ = 0.3 ($\bullet$). 
This corresponds to the time when an asteroid crosses the HZ for the first time. The statistics is made over the 10000 
asteroids and the 1$\sigma$ value is represented by error bars.}
    \label{F:time}
\end{figure}
Figure 
\ref{F:time} shows the statistical results of
the average time needed by an asteroid to become an HZc, i.e. the time it takes to reach the HZ.  This corresponds to
the time spans until the first asteroid enters the HZ. The median value and
its absolute deviation (error bars) are presented for a set of 10000 asteroids for a secondary with e$_{\scriptscriptstyle \text{b}}$ = 0.1 ($\blacksquare$) and e$_{\scriptscriptstyle \text{b}}$ = 0.3 
($\bullet$). This confirms a strong correlation between the periapsis distance and the time of first
crossing. Figure \ref{F:time} shows clearly that the average time varies from a few centuries to tens 
of thousands of
years. The closer the secondary star, the sooner asteroids can reach the HZ.

\subsection{Water transport statistics}

We now compare the water transport efficiency between binary and single star systems. For this purpose, we considered the same initial conditions for the gas giant and the asteroid belt distribution in both cases i.e. single and binary star. As the comparisons are made for the same initial conditions of the asteroids, we have to consider, for the single star system, the same size for the disk given by the binaries' characteristics. 
\begin{figure}[!h]
  \centering{
 \includegraphics[angle=-90, width=0.85\textwidth]{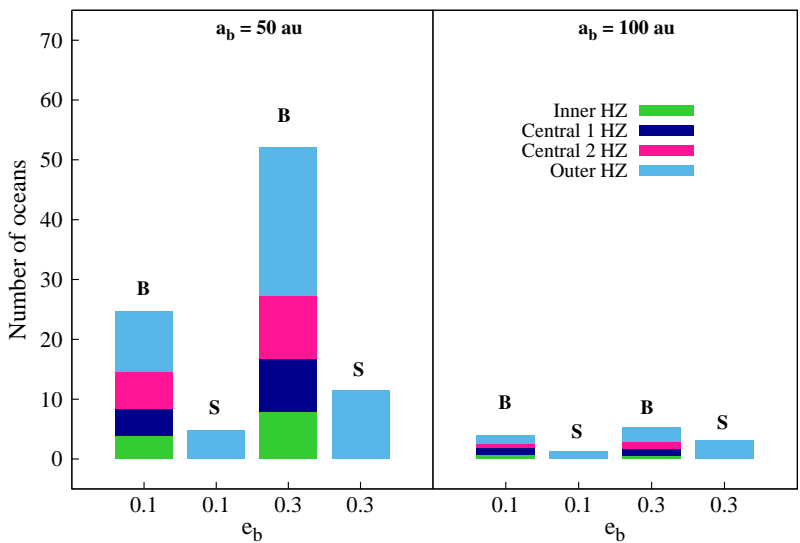}}
  \caption{Comparison of the water transport in single star (S) and binary star (B) systems. The various panels correspond to the values of $a_{\scriptscriptstyle \text{b}}$ investigated. The color code refers to different sub-rings of
the HZ (see text). We note that the two results for each panel account 
for different initial asteroid belt distributions (depending on 
$e_{\scriptscriptstyle \text{b}}$), taken as the same for (S) and (B).} 
    \label{F:oceanSS}
\end{figure}
Figure \ref{F:oceanSS} shows the total amount of water transported into the HZ (expressed in terrestrial oceans units) in various systems. The histograms marked with the letter \textit{B} refer to binary star systems, those with the letter \textit{S} to single star systems. The color code indicates the amount
of water that ended up in four equally spaced sub-rings (Inner, Central 1, Central 2, Outer) of the corresponding permanently habitable zone as defined in \cite{eggl12}. For a single star system, each sub-ring is computed using 0.950 au as
inner edge value and as outer edge value 1.676 au \citep{kasting93}. It is not 
surprising that the outer HZ is the ring with most incoming HZc. Indeed, its area is much larger than the other rings. This figure indicated also that in such single star -- giant planet, basically all the water is transported into the outer HZ 
since the perturbation is not strong enough to increase  drastically the eccentricity of an asteroid in the belt. These results show the efficiency with which a binary star  transports water into the entire HZ over a shorter
timescale compared to a single star system.

\section{Water transport: dynamics} \label{S:dynamics}

 In this section, to explain the differences in the statistical results of Sect. \ref{S:statistical}, we aim to emphasize and characterize the dynamical effects of orbital resonances (SR and MMRs), on a disk of planetesimals as well as to what extent such resonances are likely to enable the transport of water material into the HZ by icy asteroids.

\subsection{Initial modelling}

In order to highlight the influence of orbital resonances, we used a larger range of binary star configurations. The primary is still a G-type star but the secondary is either an F-, G-, K- or M-type star with mass 
$M_{\scriptscriptstyle \text{b}}$ equal to $1.3M_{\scriptscriptstyle \odot}$, $1.0M_{\scriptscriptstyle 
\odot}$, $0.7M_{\scriptscriptstyle \odot}$ and $0.4M_{\scriptscriptstyle \odot}$, respectively. The initial orbital parameters of the secondary and gas giant planet remain the same as defined in Sect. \ref{S:statistical}. In order to allow easy comparisons of the dynamics in the various systems, we consider a different distribution for the asteroid belt and therefore, we defined three different regions in our planetesimal disk:
\begin{itemize}
 \item $\mathcal{R}_{1}$: this region extends from 0.5 au to the snow line position 
at $\sim$ 2.7 au. 200 particles were initially placed in this region.
 \item $\mathcal{R}_2$: this region extends from beyond the snow line and up to the distance $a_{\scriptscriptstyle 
{\text {GG}}}- 3\,R_{\scriptscriptstyle 
{\text{H},\text{GG}}}$ $\approx$ 4.1 au. We define this region as the inner disk. As we are mainly interested in icy 
bodies that are likely to bring water to the HZ, we densified this region and 1 000 particles were distributed therein.
 \item $\mathcal{R}_3$: this region extends from $a_{\scriptscriptstyle {\text {GG}}} + 3\,R_{\scriptscriptstyle 
{\text{H},\text{GG}}}$ $\approx$ 6.3 au and up to the 
stability 
limit defined by $a_{\scriptscriptstyle {\text {c}}}$  and $\Delta\,a_{\scriptscriptstyle {\text {c}}}$. It is obvious that the size of 
the external disk will vary according to ($a_{\scriptscriptstyle {\text {b}}}$, $e_{\scriptscriptstyle {\text {b}}}$, 
$M_{\scriptscriptstyle {\text {b}}}$). The larger $a_{\scriptscriptstyle {\text {b}}}$ and the smaller 
($e_{\scriptscriptstyle {\text{b}}}$, $M_{\scriptscriptstyle {\text {b}}}$), 
the wider this region, which is called the outer disk in which 1 000 particles were placed.
\end{itemize}
For all three cases, the initial orbital separation between each particle is uniform and is defined by the ratio 
of the width of the region and the number of particles. Their initial motion is taken as nearly circular 
and planar. We also assumed that all asteroids in $\mathcal{R}_2$ and $\mathcal{R}_3,$ have equal mass and 
an initial $wmf$ of 10\%. Water mass-loss owing to ice sublimation was also taken into account during the numerical integrations. All the 
simulations are purely gravitational and we also assumed that, at this 
stage, planetary embryos have been able to form. Our simulations were performed for 50 Myr using 
the Radau integrator in the Mercury6 package \citep{chambers99}.

\subsection{Dynamics of the icy belt}

\begin{figure}[!h]
\centering{
      \includegraphics[angle=-90,width=0.49\textwidth]{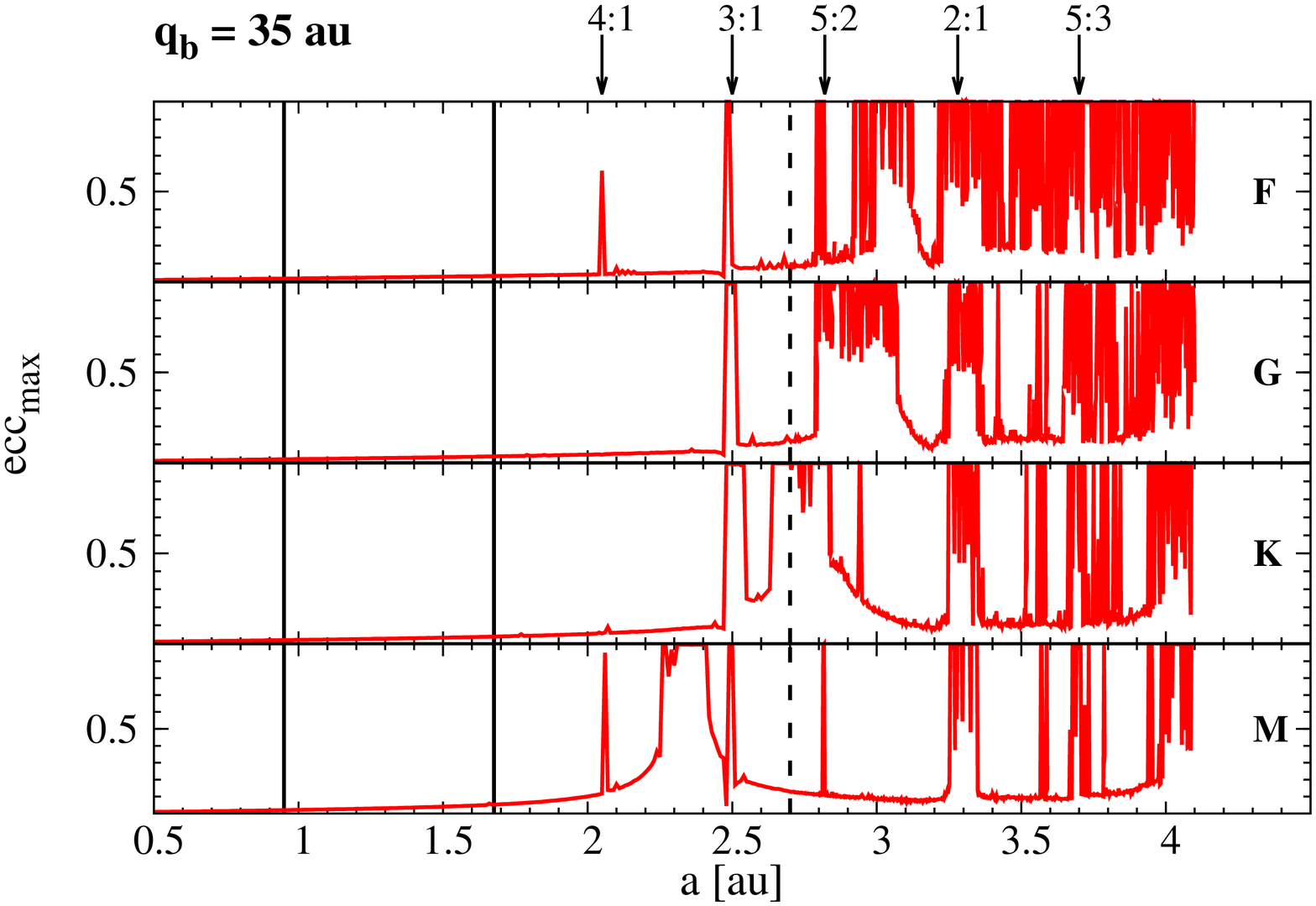} 
      \includegraphics[angle=-90,width=0.49\textwidth]{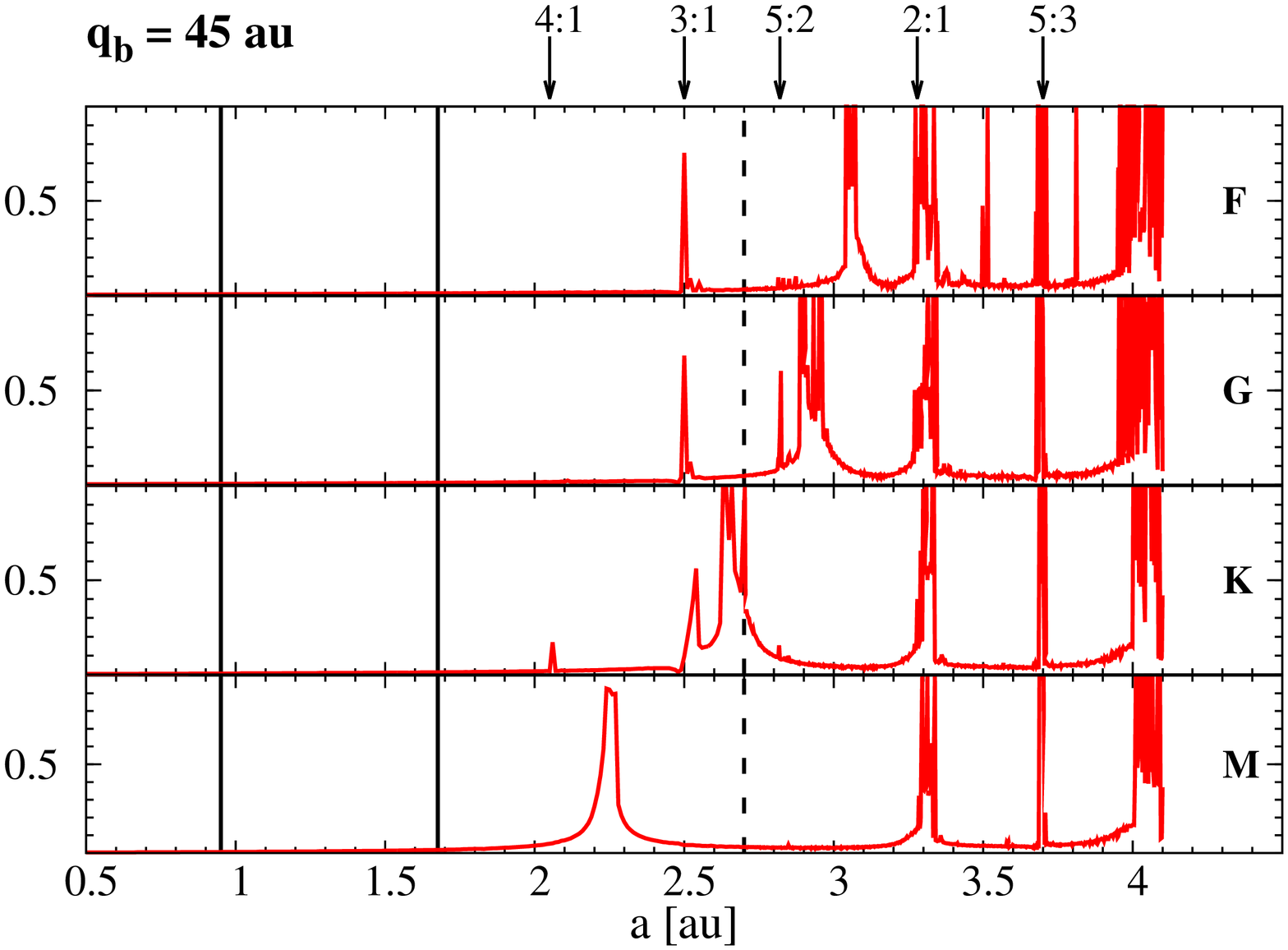} \\
      \includegraphics[angle=-90,width=0.49\textwidth]{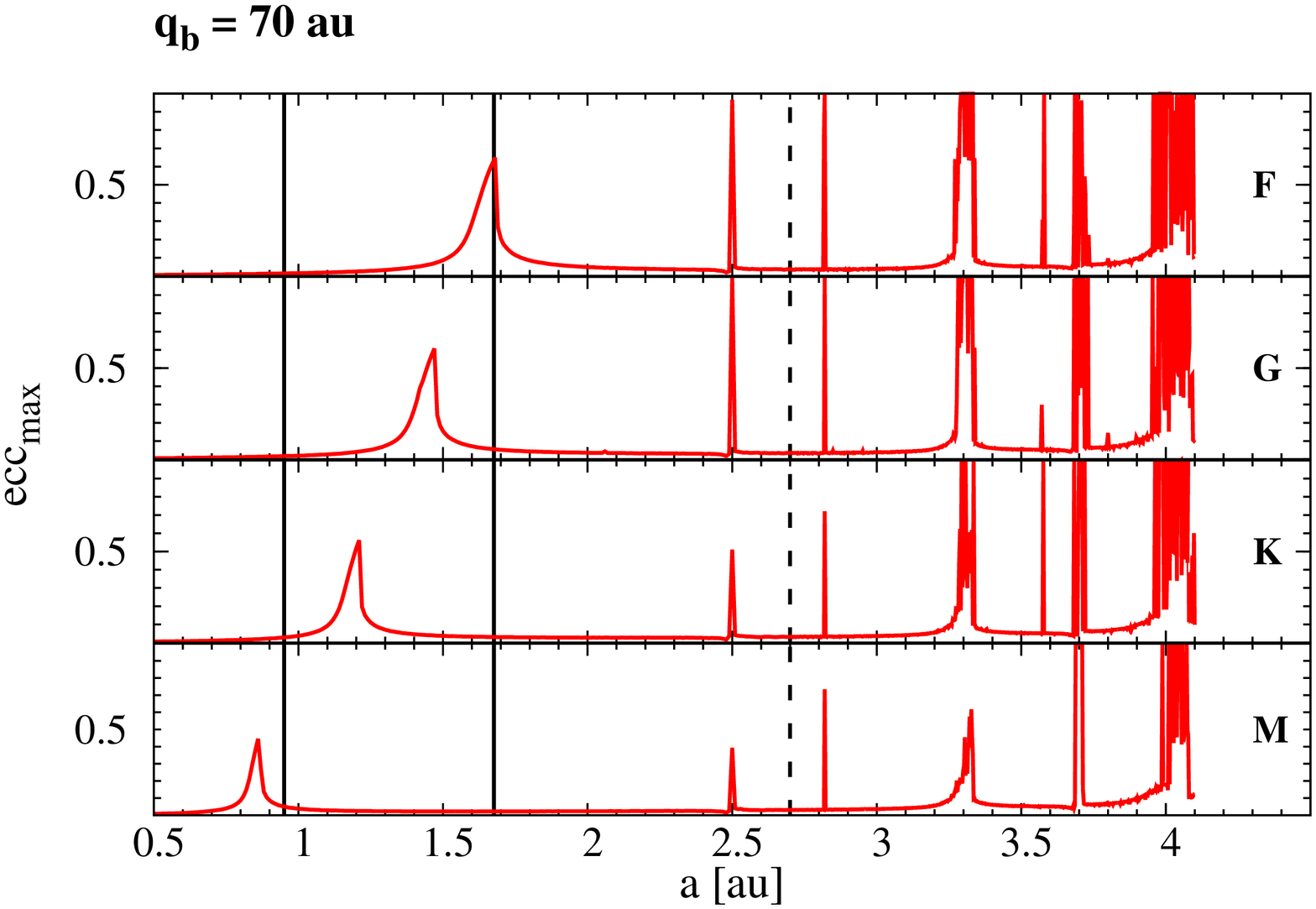} 
      \includegraphics[angle=-90,width=0.49\textwidth]{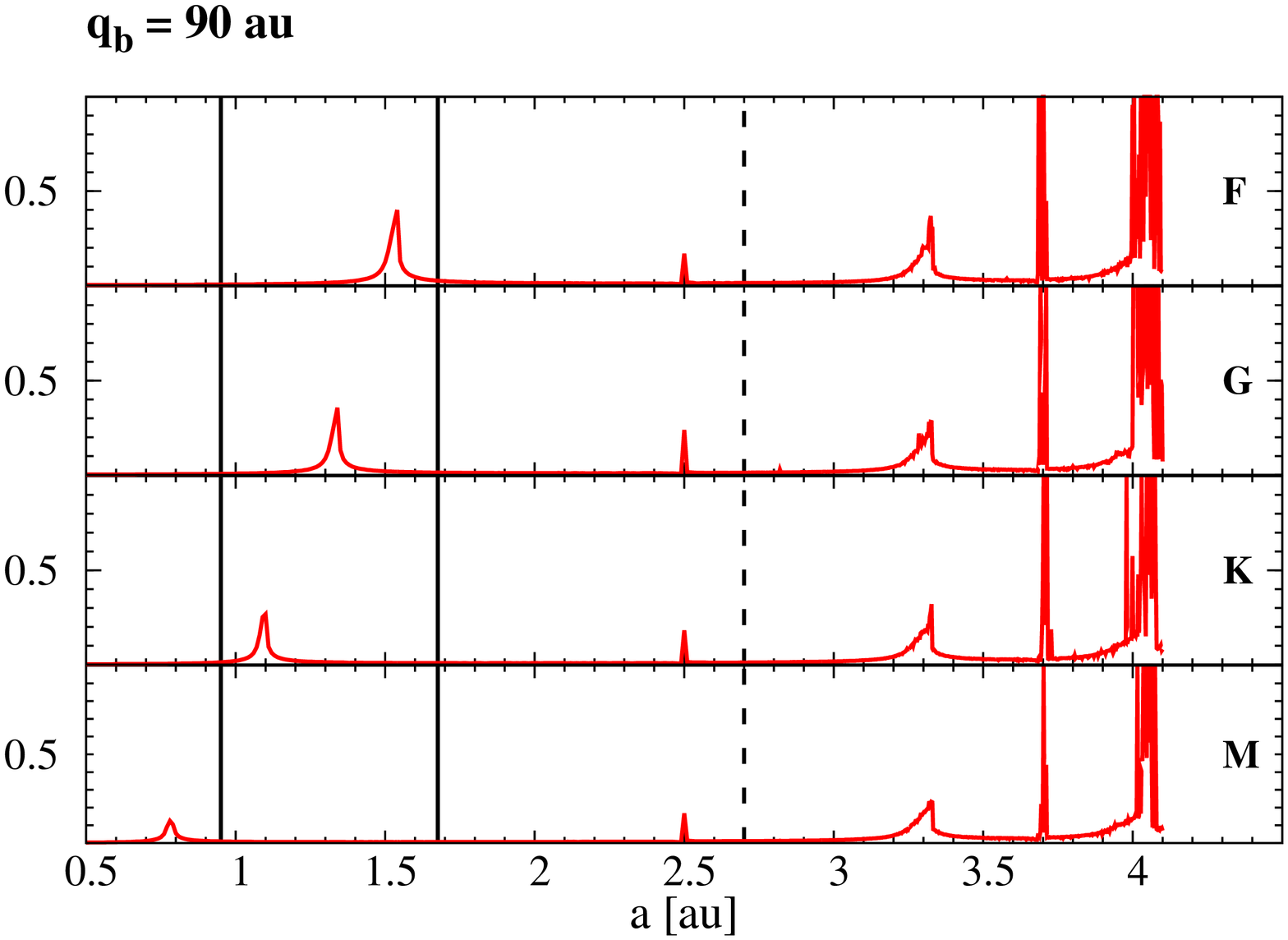} 
}
    \caption{Maximum eccentricity $ecc_{\scriptscriptstyle \text{max}}$ of planetesimals in the $\mathcal{R}_1$ and $\mathcal{R}_2$ regions, separated by the 
dashed vertical line for the snow-line position, as a function of 
their initial position, up to an intermediate integration time of 5 Myr, for different values of 
$q_{\scriptscriptstyle {\text{b}}}$. Each sub-panel refers to the 
secondary stellar type and the continuous vertical lines refer to the HZ borders. In addition, the inner main MMRs 
with 
the gas giant are indicated.}\label{F:e_inner}
\end{figure}

In Fig. \ref{F:e_inner}, we show the maximum eccentricity $ecc_{\scriptscriptstyle \text{max}}$ reached by the asteroids at different 
initial semimajor axes, in the regions $\mathcal{R}_1$ and $\mathcal{R}_2$ (separated by the vertical dashed line representing the snow-line position). The four panels correspond to the values of 
$q_{\scriptscriptstyle \text{b}}$ investigated and each sub-panel is for 
different secondary stellar types (F, G, K, and M). We can distinguish 
MMRs\footnote{Only the main ones are 
indicated} with the gas giant 
and also a secular resonance: on the bottom panels, which represent the results 
for $a_{\scriptscriptstyle \text{b}}$ = 100 au ($q_{\scriptscriptstyle \text{b}}$ = 70 au and 
$q_{\scriptscriptstyle \text{b}}$ = 90 au), we can see a spike located close to or inside the HZ\footnote{The 
borders are defined according to \cite{kopparapu13}}(continuous vertical lines) and moving outward (to larger 
semi-major 
axes) when increasing the secondary's mass. This spike represents the SR. When increasing 
$q_{\scriptscriptstyle \text{b}}$, not only does it slightly move 
inward 
but also, the maximum eccentricity reached by the particles is higher. This is because the gravitational perturbation 
from the 
secondary increases the gas giant's eccentricity. When decreasing $a_{\scriptscriptstyle \text{b}}$ to 50 au 
(top panels for 
$q_{\scriptscriptstyle \text{b}}$ = 35 au and $q_{\scriptscriptstyle \text{b}}$ = 45 au), the SR 
moves also 
outward 
and reaches the MMR region. As a consequence, the inner disk will suffer from an overlap of these orbital resonances 
that could cause a fast depletion.  However, particles inside the HZ will remain on near circular motion.
\subsection{Dynamical lifetime of asteroids near MMRs}

\begin{figure}[h!]
 \centering{\includegraphics[angle=-90,width=0.95\linewidth]{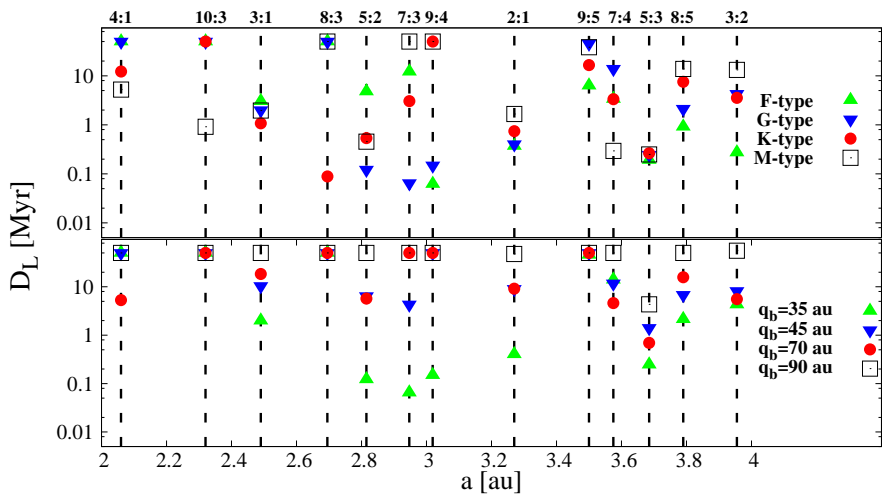}}
  \caption{Dynamical lifetime $D_{\scriptscriptstyle \text{L}}$ of test particles in $\mathcal{R}_1$ and $\mathcal{R}_2$ regions expressed in Myr. Top 
panel: influence of the 
secondary's mass when $q_{\scriptscriptstyle {\text{b}}}$ = 35 au. Bottom panel: 
influence of the secondary's periapsis distance $q_{\scriptscriptstyle \text{b}}$  assuming the secondary as a G-type 
star.}
     \label{F:MMR}
\end{figure}

We investigate in detail the dynamical lifetime of massless particles $D_{\scriptscriptstyle \text{L}}$ which are initially close or inside 
internal MMRs. 
They occur when the orbital periods of the gas giant and the particle are in commensurability, such as
\begin{equation*}
 \displaystyle a _{\scriptscriptstyle \text{n}} = \left (\frac{\text{p}}{\text{q}} \right)^{\scriptscriptstyle 
{2/3}} a_{\scriptscriptstyle {\text{GG}}}\,\mbox{,}~~~\mbox{p and q} \in \mathbb{N} \displaystyle \left \{ 
\begin{array}{r}
                                              \text{p}>\text{q} ~~\mbox{if} ~~\mbox{} a_{\scriptscriptstyle 
\text{n}} < a_{\scriptscriptstyle \text{GG}} \\
                                              \text{p}<\text{q} ~~\mbox{if} ~~\mbox{} a_{\scriptscriptstyle 
\text{n}} > a_{\scriptscriptstyle \text{GG}} 
                                            \end{array} \right.,
\end{equation*}
where $a_{\scriptscriptstyle \text{n}}$ is the position of the nominal resonance. As we aim to correlate $D_{\scriptscriptstyle \text{L}}$ with the binary star characteristics, we preferred to do a separate analysis to ensure that each MMR contains the same number of particles. We limited this study to resonances with integers p and q $\le$ 10. In each MMR, we uniformly distributed 25 particles initially on 
circular and planar orbits. In addition, as suggested by the studies of \cite{holman99} and \cite{pilat02}, each particle is cloned into four starting 
points with mean anomalies 0$^\circ$, 90$^\circ$, 180$^\circ$ and 270$^\circ$, since it is 
well known that the starting position plays an important role for the dynamical 
behaviour in MMRs. This accounts for 100 particles in each MMR. Each 
system was integrated for 50 Myr. We consider a particle as leaving its initial location inside a specific MMR when its dynamical evolution leads to a collision with either one of the stars or the gas giant. Finally, we defined $D_{\scriptscriptstyle \text{L}}$ as the time required for 50\% of the population to leave a specific MMR \citep{gladman97}. In Fig. \ref{F:MMR}, we show the dynamical lifetime in Myr of particles near the internal MMRs. On the top panel of this figure, the influence of $M_{\scriptscriptstyle {\text{b}}}$ is shown for a certain periapsis distance $q_{\scriptscriptstyle {\text{b}}}$ = 35 au as for this particular value, the SR overlaps with the MMRs in $\mathcal{R}_2$. The bottom panel summarizes the results for a certain mass of the secondary star (i.e. G-type star) and different periapsis distances of this star.

We can see that prior to the 8:3 MMR, the border between 
icy and rocky asteroids, a secondary M-type will favour chaos inside the rocky bodies region located in 
$\mathcal{R}_1$. This is not surprising since Fig. \ref{F:e_inner} clearly shows the SR overlapping 
with MMRs located inside the snow-line at 2.7 au. Beyond this limit, a higher 
value of $M_{\scriptscriptstyle {\text{b}}}$ leads to a lower $D_{\scriptscriptstyle \text{L}}$ -- values can reach 0.1 Myr -- since the SR moves outward. From the bottom panel, one can 
recognize that the lower $q_{\scriptscriptstyle 
\text{b}}$, the 
lower $D_{\scriptscriptstyle \text{L}}$. Some MMRs 
can be quickly emptied within 0.1 Myr. With these tests, we highlight that in binary star 
systems, $D_{\scriptscriptstyle \text{L}}$ of particles initially orbiting inside or close to MMRs can be variable according to the location of the 
SR. Provided that particles can reach the HZ region before colliding with one of the massive bodies 
(i.e. the stars or the gas giant) or before being ejected from the system, 
they could rapidly cause an early bombardment on any embryo or planet moving in the HZ.

\subsection{Consequences for the water transport}
\begin{figure}[!h]
\centering{
  \includegraphics[angle=-90,width=\textwidth]{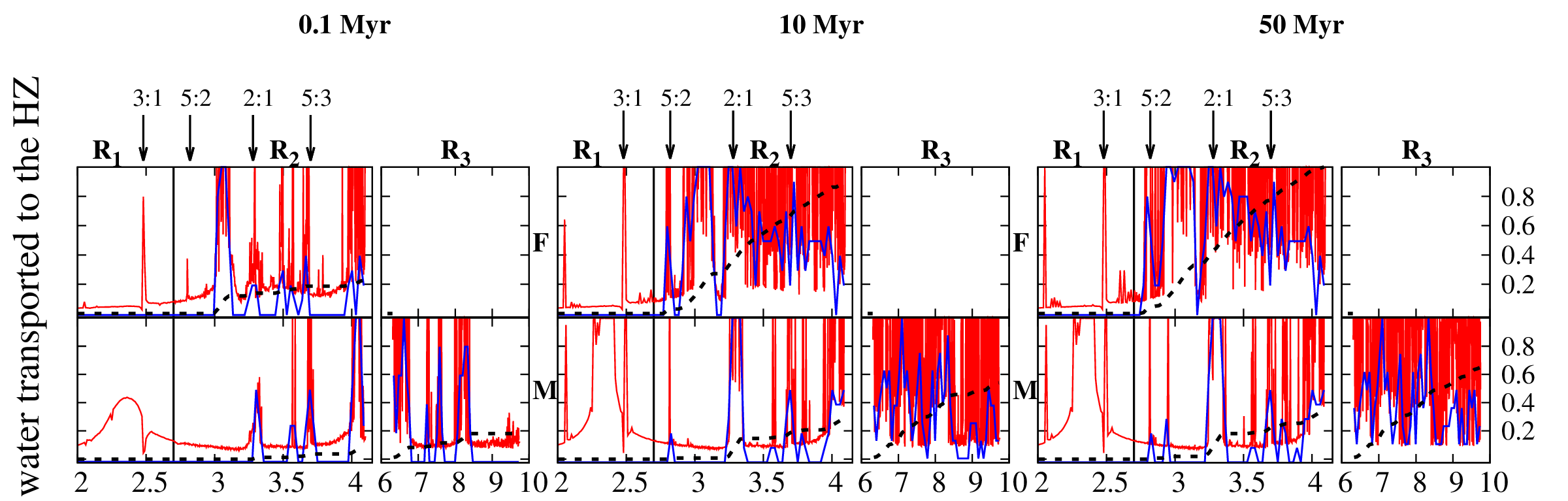}\\[1.2ex]
  \includegraphics[angle=-90,width=\textwidth]{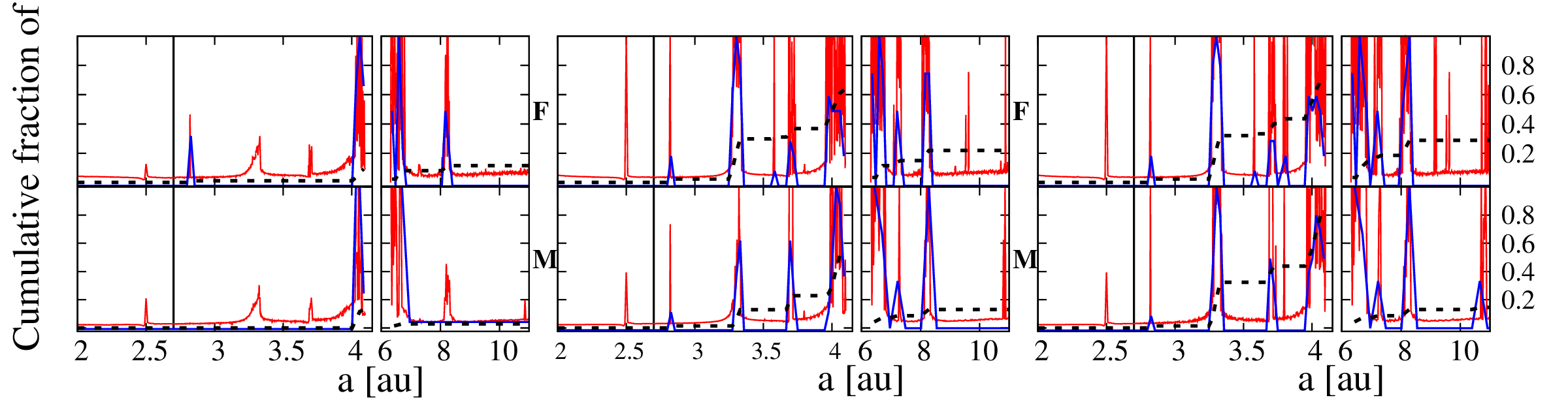}\\
}
  \caption{Represented on the y-axis (left and right axes have the same scale), $ecc_{\scriptscriptstyle \text{max}}$ (red line) together with the cumulative fraction of water (dashed black line) brought to the HZ, with respect to the initial location of small bodies in $\mathcal{R}_2$ and $\mathcal{R}_3$. The top panel is for a secondary star at $q_{\scriptscriptstyle \text{b}}$ = 35 au and the bottom panel is for $q_{\scriptscriptstyle \text{b}}$ = 70 au. In this figure, is also represented the normalized HZc distribution (blue line) from 0.1 Myr (left) up to 50 Myr (right). The vertical black line refers to the position of the snow line.}
     \label{F:e_HZc}
\end{figure}

Here, we compare the flux of icy particles from $\mathcal{R}_2$ and 
$\mathcal{R}_3$ towards the HZ (see also Figs. 6 and 7 in \cite{bancelin16}). On the $y$-axis in Fig. \ref{F:e_HZc}, we represent the evolution of $ecc_{\scriptscriptstyle \text{max}}$ (red line) of particles inside $\mathcal{R}_1$, $\mathcal{R}_2$ (as already drawn in Fig. \ref{F:e_inner}) and $\mathcal{R}_3$. The top panel corresponds to a secondary star at $q_{\scriptscriptstyle \text{b}}$ = 35 au and the bottom panel is for $q_{\scriptscriptstyle \text{b}}$ = 70 au, each sub-panel corresponding either to a F- or M-type secondary star. 

From left to right, 
each figure is for a different period of integration (0.1 Myr, 10 Myr and 50 
Myr). We also show the normalized HZc distribution (blue line) calculated regarding the total number of HZc produced by the corresponding systems for each period of integration time. For all  cases considered in the top panels ($q_{\scriptscriptstyle \text{b}}$ = 35 au), within 10 Myr, the 2:1 MMR located at 
$\sim$ 3.28 au and the SR, when it lies beyond the snow line, are the 
primary sources of HZc in the inner disk.  In addition, the external disk can 
produce a non-negligible or equivalent number of HZc, compared to the inner disk. On the same figure, the $y$-axis also corresponds to the cumulative fraction of water (dashed black line) brought by the HZc. This fraction is determined with respect to the final amount of transported water from $\mathcal{R}_2$ and $\mathcal{R}_3$ within 
50 Myr. For the masses investigated at $q_{\scriptscriptstyle \text{b}}$ = 35 au, the results exhibit the same trend: the quantity of incoming water inside the HZ drastically increases when particles orbit initially inside the SR and the 2:1 MMR.

For $q_{\scriptscriptstyle \text{b}}$ = 70 au (bottom panels), contrary to the previous case, the SR does not contribute at all to bearing water material into the HZ since it lies in this region (see Fig. \ref{F:e_inner}). We show that the two main sources of HZc in $\mathcal{R}_2$ are the 2:1 and 5:3 MMR. The contribution of $\mathcal{R}_3$ is more negligible than in the previous case since its size is more extended and weakly perturbed.

\section{Influence of orbital resonances on the water transport} \label{S:EP}

 The results of Sect. \ref{S:dynamics} revealed how secular and mean motion resonances can affect the dynamics of an asteroid belt. However, we also showed that for a given binary star -- gas giant configuration, particles initially orbiting inside the HZ or beyond the snow-line can show different dynamical outcomes depending on the location of the SR i.e. inside or outside the HZ.
 
The aim of this section is to highlight in detail the influence of orbital resonances on the water transport by icy asteroids from a planetesimal disk, initially beyond the snow-line, to embryos-to-planet-sized bodies (EPs), initially inside the HZ. We mainly focused on the following scenarios:
\begin{itemize}
	\item [a)] eccentric planetary motion inside the HZ (induced by an SR at $\sim$ 1.0 au and/or MMRs) and an asteroid belt perturbed by MMRs
	\item [b)] nearly circular planetary orbits inside the HZ and an asteroid belt 
	perturbed by an SR and MMRs.
\end{itemize}

\subsection{Initial modelling}

As in Sect. \ref{S:statistical}, we investigate only binary star systems with two G-type stars with same initial orbital parameters and also hosting a gas giant planet. However, as pointed out by \cite{pilat15}, the location of the SR depends both on the orbital elements of the secondary and the gas giant. Since we fixed the binary's semi-major axis to 50 and 100 au for our numerical study, the position of the gas giant $a_{\scriptscriptstyle \text{GG}}$ changes for the different systems depending on the investigated scenarios described above. With the conditions that the SR should either be beyond 2.7 au or around 1.0 au, we determine the according semi-major axis of the giant planet using the semi-analytical method of \cite{pilat15} and \cite{bazso15}. We summarize the values of $a_{\scriptscriptstyle \text{GG}}$ in Tab. \ref{T:agg}.
\begin{table}[!h]
	\begin{center}
		\caption{Gas giant's semi-major axis $a_{\scriptscriptstyle \text{GG}}$ according to the binary's periapsis 
			distance $q_{\scriptscriptstyle \text{b}}$. Second and third columns are values 
			of $a_{\scriptscriptstyle \text{GG}}$ 
			such as the secular resonance (SR) lies inside the HZ or inside the asteroid belt, 
			respectively.}
		\label{T:agg}
		\begin{tabular}{ccc}
			\hline
			$q_{\scriptscriptstyle \text{b}}$ [au] & $a_{\scriptscriptstyle \text{GG}}$ [au] & $a_{\scriptscriptstyle \text{GG}}$ 
			[au] \\
			& (SR $\in$ HZ) & (SR $\notin$ HZ)\\
			\hline
			35 & 3.00 & 5.2 \cr
			\hline
			45 & 3.10 & 5.2 \cr
			\hline 
			70 & 4.75 & 7.2\cr
			\hline
			90 & 4.5 & 7.2 \cr
			
			\hline
		\end{tabular}
	\end{center}
\end{table}
We uniformly distribute 1000 icy Ceres-like asteroids at initial semi-major axis $a$ into three regions as defined as follow:
\begin{itemize}
	\item $\mathcal{R}_{1}^{\prime}$ is for 2.0 $\le$ $a$ $\le$ 2.7 au,
	\item $\mathcal{R}_{2}^{\prime}$ is for 2.7 $\le$ $a$ $\le$ $a_{\scriptscriptstyle {\text {GG}}} - 3\,R_{\scriptscriptstyle {\text{H},\text{GG}}}$,
	\item $\mathcal{R}_{3}^{\prime}$ is for $a$ $\ge$ $a_{\scriptscriptstyle {\text {GG}}} +
	3\,R_{\scriptscriptstyle {\text{H},\text{GG}}}$
\end{itemize}
and each asteroid is assigned a $wmf$ using a linear approximation and with borders defined in the following way: 
\begin{itemize}
	\item $a \in\,\mathcal{R}_{1}^{\prime}$, 1 $\le$ $wmf$ $\le$ 10\%
	\item $a \in\,\mathcal{R}_{2}^{\prime}$, 10 $\le$ $wmf$ $\le$ 15\%
	\item $a \in\,\mathcal{R}_{3}^{\prime}$, $wmf = 20\%$
\end{itemize}

Contrary to the previous sections, as the distribution of $a$ within the disk is different for the various binaries investigated (as it depends on the location of the gas giant), the total amount of water $\mathcal{W}_{\scriptscriptstyle \text{TOT}}$ (expressed in Earth-ocean unit) contained in the disk varies as shown in Tab. \ref{T:tot_water}.

\begin{table}[!h]
	\begin{center}
		\caption{Total amount of water $\mathcal{W}_{\scriptscriptstyle \text{TOT}}$ (expressed in Earth-ocean unit) contained in the icy asteroids belt.}
		\label{T:tot_water}
		\begin{tabular}{ccc}
			\hline
			$q_{\scriptscriptstyle \text{b}}$ [au] & $\mathcal{W}_{\scriptscriptstyle \text{TOT}}$ [Earth-ocean] &  $\mathcal{W}_{\scriptscriptstyle \text{TOT}}$ [Earth-ocean] \\
			& (SR $\in$ HZ) & (SR $\notin$ HZ)\\
			\hline
			35 & 58.6 & 78.2 \cr
			\hline
			45 & 60.2 & 78.2 \cr
			\hline 
			70 & 75.6 & 81.5\cr
			\hline
			90 & 76.7 & 81.5 \cr
			
			\hline
		\end{tabular}
	\end{center}
\end{table}
\subsection{Collision modelling}

For both scenarios, i.e. the SR either inside the HZ or inside the asteroid belt, we performed two simulations for 100 Myr using the Radau integrator in order to simulate impacts between icy objects initially located in the asteroid belt and 
EPs initially inside the HZ. They are either Moon- and Mars-sized (embryos) or Earth-sized (planets). Considering the gravitational perturbations of the stars and the gas giant, we integrated separately:
\begin{itemize}
	\item the asteroid belt to assess the orbital distribution of asteroids crossing the (x,y) plane, with a distance $r_{\scriptscriptstyle \text{A}} < 2.0$ au 
	to the primary star.
	\item EPs at initial location $a_{\scriptscriptstyle {\text{EP}}}$ to assess the evolution of their eccentricity with time. They are uniformly distributed over 48 positions within the HZ (with borders defined according to \cite{kopparapu13}) and initially move in circular orbits.
\end{itemize}

We combined the results of these two integrations to analytically compute the minimum orbital intersection 
distance (MOID) 
\citep{sitarski68}. The MOID corresponds to the closest distance between two 
keplerian orbits, regardless their real position on their respective 
trajectories. 
We define a collision if the MOID is comparable to the EP's radius $R_{\scriptscriptstyle \text{EP}}$. For the collision assessment, we use the following  five-steps algorithm:
\begin{itemize}
	\item [(1)] For a given position $a_{\scriptscriptstyle \text{EP}}$, we check if $q_{\scriptscriptstyle \text{EP}}$ 
	$\le$ 
	$r_{\scriptscriptstyle \text{A}}$ $\le$ $Q_{\scriptscriptstyle \text{EP}}$ where $q_{\scriptscriptstyle \text{EP}}$ and 
	$Q_{\scriptscriptstyle \text{EP}}$ are the periapsis and apoapsis 
	distances of the EPs respectively which are defined according to 
	$e_{\scriptscriptstyle \text{EP}}$ = $e_{\scriptscriptstyle \text{max}} (t)$ with 
	$e_{\scriptscriptstyle \text{max}} (t)$ the maximum of the EP's eccentricity for 
	different 
	period of integration. If this condition is fulfilled, then we go to the next step.
	
	\item [(2)] In reality, $e_{\scriptscriptstyle \text{EP}}$ has periodic variations between 0 (its initial value) and 
	$e_{\scriptscriptstyle 
		\text{max}} (t)$. Thus, we define the function $Y(e_{\scriptscriptstyle \text{EP}})$ = 
	$d_{\scriptscriptstyle \text{min}} 
	(e_{\scriptscriptstyle \text{EP}})$ - $R_{\scriptscriptstyle \text{EP}}$ where $d_{\scriptscriptstyle \text{min}}$ is 
	the MOID. According to the sign of the product $Y(e_{\scriptscriptstyle \text{EP}} = 0)$ $\times$ 
	$Y(e_{\scriptscriptstyle \text{EP}} = e_{\scriptscriptstyle \text{max}})$ we use different procedures: if the sign 
	is negative then we use a 
	regula falsi procedure in order to find the value of $e_{\scriptscriptstyle \text{EP}}$ giving $Y(e_{\scriptscriptstyle 
		\text{EP}}) \approx 0$. If the sign is positive, then we use a dichotomy 
	procedure to find a value 
	of $e_{\scriptscriptstyle \text{EP}}$ leading to an impact.
	
	\item [(3)] When a collision is found, i.e. if $d_{\scriptscriptstyle {\text{min}}}\,\le\,R_{\scriptscriptstyle 
		{\text{EP}}}$ we also derive, from the MOID, the true anomalies of the EP and 
	the asteroid in order to 
	compute the relative impact velocity and impact angle of the asteroid.
	
	\item [(4)] If the previous condition is fulfilled, then we define:
	\begin{equation*}
	\left\{ 
	\begin{array}{l}
	\mathcal{W}_{\scriptscriptstyle \text{k}}(a_{\scriptscriptstyle \text{EP}}, t) = \displaystyle \frac{M_{\scriptscriptstyle \text{CERES}}\, \times wmf_{\scriptscriptstyle \text{k}}}{M_{\scriptscriptstyle \text{H}_{\scriptscriptstyle 2} \text{O}}} \\
	
	\displaystyle \widetilde{\mathcal{W}}_{\scriptscriptstyle \text{k}}(a_{\scriptscriptstyle \text{EP}}, t) = \frac{1}{N_{\scriptscriptstyle \text{k}}}  \displaystyle \sum_{k=1}^{N_{\scriptscriptstyle \text{k}}} \frac{M_{\scriptscriptstyle \text{CERES}}\, \times wmf_{\scriptscriptstyle \text{k}}\times\,\left( 1-\omega_{\scriptscriptstyle \text{c}} (a_{\scriptscriptstyle \text{EP}}) \right)}{M_{\scriptscriptstyle \text{H}_{\scriptscriptstyle 2} \text{O}}}
	\end{array}
	\right.
	\end{equation*}
	
 as the quantity of water (in Earth-ocean unit) delivered by an asteroid number $k$ (with water mass fraction $wmf_{\scriptscriptstyle \text{k}}$), at an intermediate integration time $t$, to the EP initially at a semi-major axis $a_{\scriptscriptstyle \text{EP}}$. Here, $M_{\scriptscriptstyle \text{H}_{\scriptscriptstyle 2} \text{O}} = 1.5\,\times 10^{21}$ kg of H$_2$O is the mass of one Earth-ocean. The first term $\mathcal{W}_{\scriptscriptstyle \text{k}}$ assumes a merging approach in which the whole water content of the asteroid is delivered to the EP without assuming any water loss processes. The second term $\displaystyle \widetilde{\mathcal{W}}_{\scriptscriptstyle \text{k}}$ takes into account a water loss factor $\omega_{\scriptscriptstyle \text{c}}$ induced by a water loss mechanism. Here, $N_{\scriptscriptstyle \text{k}}$ is the number of possible collisions of the asteroid $k$.
	
	\item [(5)] At the end of the procedure, we can derive the total fraction of water delivered to the EP: $\mathcal{W}_{\scriptscriptstyle \text{EP}} (a_{\scriptscriptstyle \text{EP}},t) = \displaystyle \frac{1}{\mathcal{W}_{\scriptscriptstyle \text{TOT}}} \sum_{k=1}^{n_{\scriptscriptstyle \text{i}}}\mathcal{W}_{\scriptscriptstyle \text{k}}(a_{\scriptscriptstyle \text{EP}},t)$ and $\mathcal{\widetilde{W}}_{\scriptscriptstyle \text{EP}} (a_{\scriptscriptstyle \text{EP}},t) = \displaystyle \frac{1}{\mathcal{W}_{\scriptscriptstyle \text{TOT}}} \sum_{k=1}^{n_{\scriptscriptstyle \text{i}}}\mathcal{\widetilde{W}}_{\scriptscriptstyle \text{k}}(a_{\scriptscriptstyle \text{EP}},t)$ where $n_{\scriptscriptstyle \text{i}}$ is the number of impactors. We also compute the median values of impact velocities $\overline{v}_{\scriptscriptstyle \text{i}}(a_{\scriptscriptstyle \text{EP}})$ and angles $\overline{\theta}_{\scriptscriptstyle \text{i}}(a_{\scriptscriptstyle \text{EP}})$ according to the total number of possible collisions found.

\end{itemize}

\subsection{Dynamics statistics}

\begin{figure}[!h]
 \centering{\includegraphics[angle=-90,width=\linewidth]{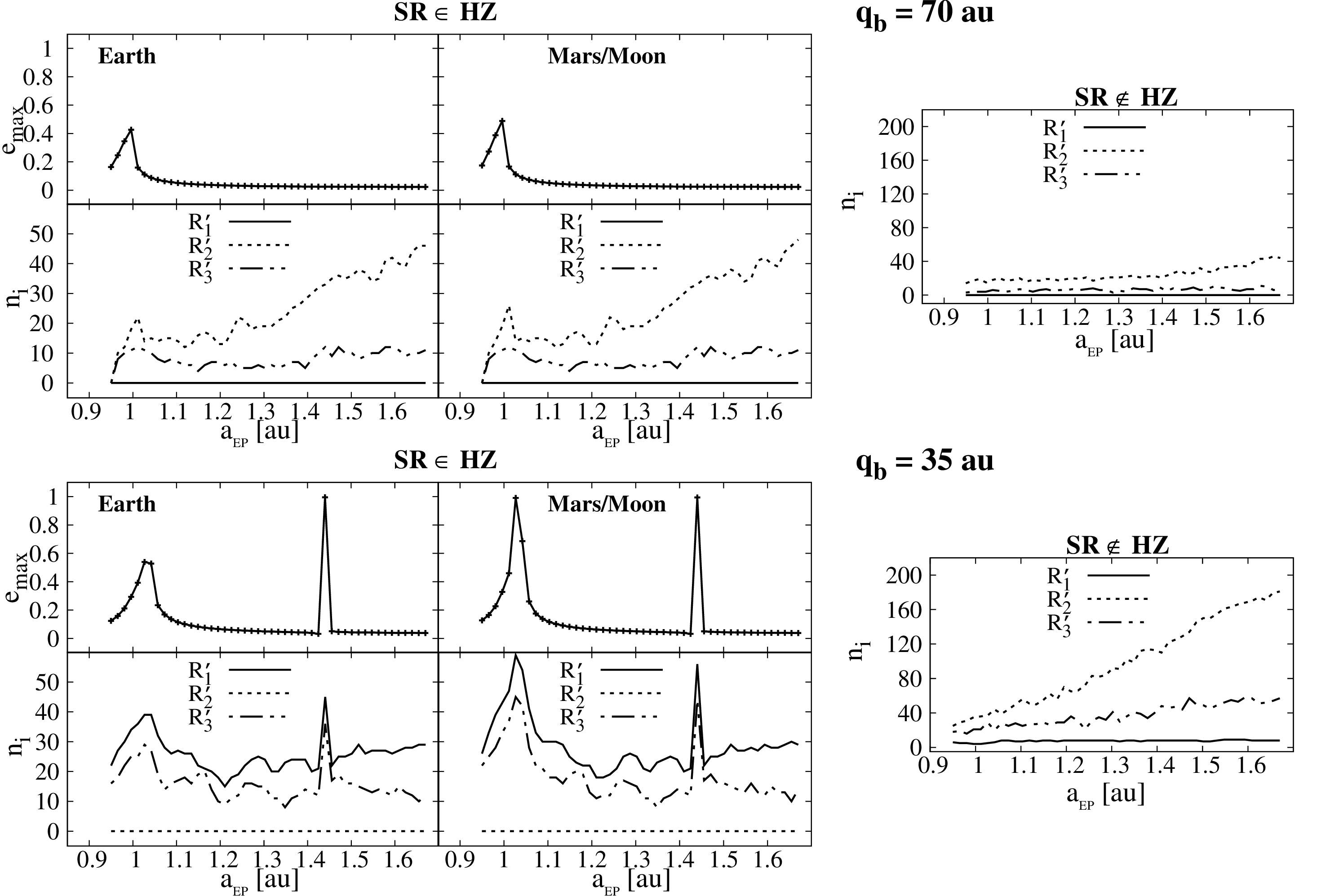}}
  \caption{$Left$ and $right$ panels: results for SR $\in$ HZ and SR $\notin$ HZ, respectively for $q_{\scriptscriptstyle \text{b}}$ = 35 and 70 au, $bottom$ and $top$ panels, respectively. For a given value of $q_{\scriptscriptstyle \text{b}}$, $e_{\scriptscriptstyle \text{max}}$ and $n_{\scriptscriptstyle \text{i}}$ are represented on the $top$ and $bottom$ sub-panels for the case of an Earth- ($left$ sub-panel) and Mars/Moon-sized objects ($right$ sub-panel). For SR $\notin$ HZ, as the eccentricity evolution is the same for each EP, only the $n_{\scriptscriptstyle \text{i}}$ of one the EPs is displayed.}
     \label{F:dyn_imp}
\end{figure}

In Fig. \ref{F:dyn_imp}, we compare results obtained when an SR is inside and outside the HZ (left and right panels, respectively) for a binary at $a_{\scriptscriptstyle \text{b}}$ = 50 (bottom panels) and 100 au (top panels) with $e_{\scriptscriptstyle \text{b}}$ = 0.3 (see also Figs. 2 and 3 in \cite{bancelin17} for more configurations). In addition, for the SR inside the HZ, the top sub-panels represent the maximum eccentricity $e_{\scriptscriptstyle \text{max}}$ of the EPs and the bottom sub-panels are for the number of impactors $n_{\scriptscriptstyle \text{i}}$. The results for an Earth-  are shown on the left sub-panels and for Mars/Moon-sized objects on the right sub-panels. For the SR inside the belt, only $n_{\scriptscriptstyle \text{i}}$ is displayed as the EPs' orbit remain almost circular during the integration time, regardless the size of the EP.  One can recognize in the left panels the SR around 1.0 au causing a relatively high eccentric motion with decreasing $q_{\scriptscriptstyle \text{b}}$ and $M_{\scriptscriptstyle \text{EP}}$. Due to the proximity of the gas giant, we can also notice, for $q_{\scriptscriptstyle \text{b}}$ = 35 au, the presence of the 3:1 MMR at 1.44 au also causing high eccentric motions in this area. 
When the SR is inside the HZ (left panels), we can notice that EPs can collide with more asteroids than if they were orbiting outside the orbital resonances. Both regions $\mathcal{R}_{1}^{\prime}$ and $\mathcal{R}_{3}^{\prime}$ can be water sources for EPs within the entire HZ and mainly at $\sim$ 1.0 au. However, an SR located inside the belt (right panels) can boost the number of impactors on the EP's surface and this can lead up to 200 impactors from region $\mathcal{R}_2^{\prime}$ for $q_{\scriptscriptstyle \text{b}}$ = 35 au. One can notice for the case the SR is not in the HZ that most of the impactors are located mainly in the outer HZ border. This was already pointed out in \cite{bancelin15}. Therefore, only a small fraction of $n_{\scriptscriptstyle \text{i}}$ can impact at $\sim$ 1.0 au but this number is comparable to the case of the SR located inside the HZ.

\subsection{Collision parameters statistics}
\begin{figure}[!h]
	\centering{
		\includegraphics[angle=-90,width=\textwidth]{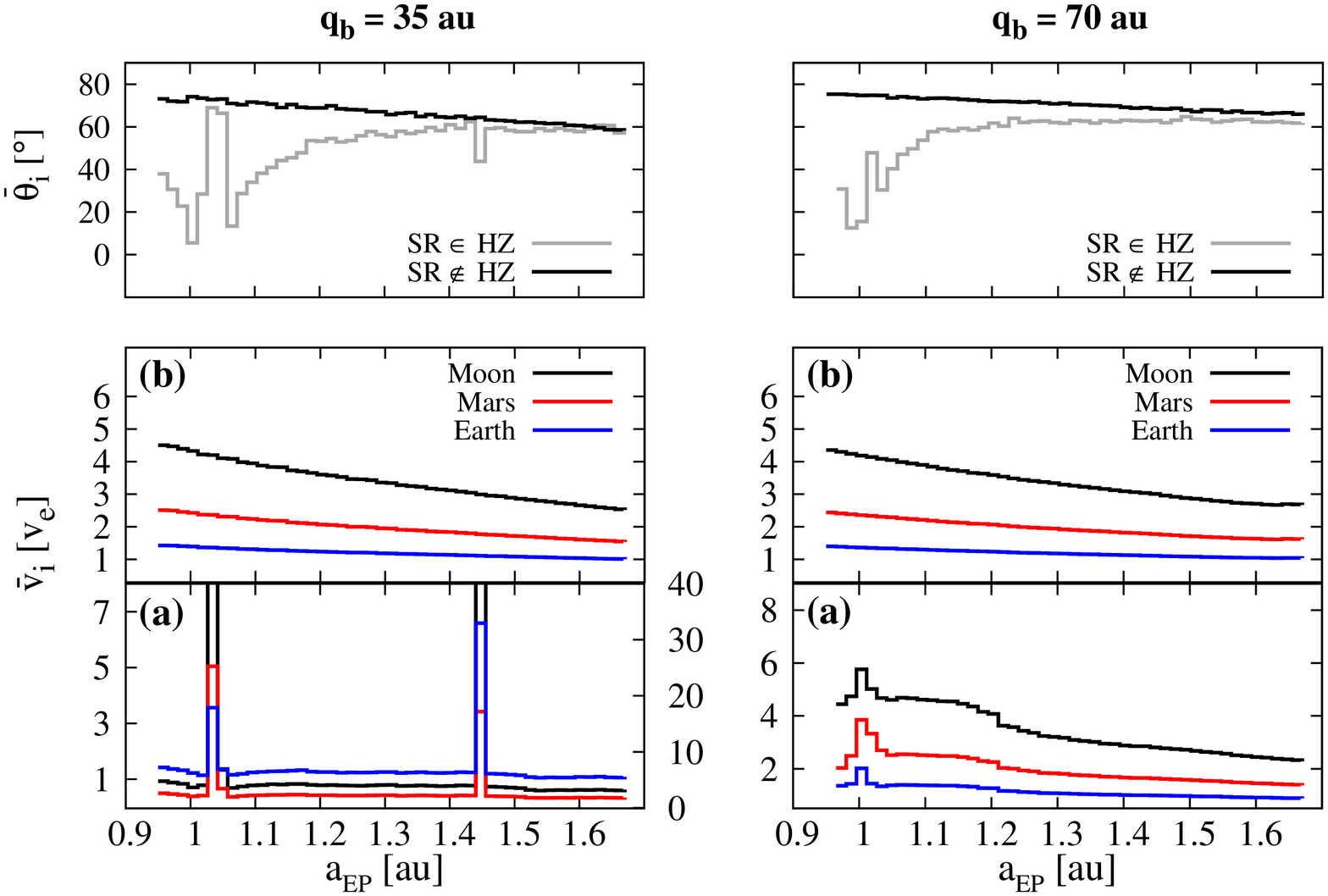}}
	\caption{Impact angles $\overline{\theta}_{\scriptscriptstyle \text{i}}$ velocities $\overline{v}_{\scriptscriptstyle \text{i}}$ (\textit{top} and \textit{bottom} panels respectively) with respect to $a_{\scriptscriptstyle \text{EP}}$ and for $q_{\scriptscriptstyle \text{b}}$ = 35 and 70 au, $left$ and $right$ panels, respectively. Colors in the \textit{bottom} panels are for impact velocities at the surface of Earth, Mars and Moon objects (solid blue, red and black lines respectively) depending on if SR lies inside or outside the HZ (panels \textit{a} and \textit{b} respectively). For $q_{\scriptscriptstyle \text{b}}$ = 35 au, values of $\overline{v}_{\scriptscriptstyle \text{i}}$ for Moon/Mars have to be read on the right y-axis as the values are much larger than the Earth (left y-axis). Colors in the \textit{top} panel is for SR inside the HZ (solid grey line) or inside the belt (solid black line).}\label{F:impact}
	
\end{figure}

In Fig. \ref{F:impact} (see also Fig. 4 in \cite{bancelin17} for more configurations), we represent the impact angles and velocities distribution $\overline{\theta}_{\scriptscriptstyle \text{i}}$ and $\overline{v}_{\scriptscriptstyle \text{i}}$ (top and bottom panels respectively). Results are compared for both locations of the SR: inside the HZ (grey solid line for $\overline{\theta}_{\scriptscriptstyle \text{i}}$ and panels (a) for $\overline{v}_{\scriptscriptstyle \text{i}}$) and outside the HZ (black solid line for $\overline{\theta}_{\scriptscriptstyle \text{i}}$ and panels (b) for $\overline{v}_{\scriptscriptstyle \text{i}}$). In the bottom panels, we distinguished impact velocities on Earth-, Mars- and Moon-sized objects (solid blue, red and black lines respectively).
For both cases (SR inside and outside the HZ), the impact velocity distribution 
inside the HZ shows different modes depending $a_{\scriptscriptstyle \text{EP}}$: outside the orbital resonances, statistically, we have for an 
Earth-sized $1.0 < \overline{v}_{\scriptscriptstyle \text{i}} < 1.5\,v_{\scriptscriptstyle \text{e}}$, for a Mars-sized $1.5< \overline{v}_{\scriptscriptstyle \text{i}} < 
2.5\,v_{\scriptscriptstyle \text{e}}$ and for a Moon-sized $2.5 < \overline{v}_{\scriptscriptstyle \text{i}} < 6.0\,v_{\scriptscriptstyle \text{e}}$ (all 
expressed in escape velocity 
$v_{\scriptscriptstyle \text{e}}$ units of the respective EP). However, near an orbital resonance, 
$\overline{v}_{\scriptscriptstyle \text{i}}$ can be slightly higher as seen in Fig. 
\ref{F:impact}. For instance, for $q_{\scriptscriptstyle \text{b}}$ = 35 au (for this particular system, one should read results for an Earth on the left $y$-axis and for a Mars/Moon on the right $y$-axis), $\overline{v}_{\scriptscriptstyle \text{i}}$ can reach up to 6.0$\,v_{\scriptscriptstyle \text{e}}$ for an Earth-, 25$\,v_{\scriptscriptstyle \text{e}}$ for a Mars- and beyond 
40$\,v_{\scriptscriptstyle \text{e}}$ for a Moon-sized body. In a similar way, $\overline{\theta}_{\scriptscriptstyle \text{i}}$ has high variations between 20$^\circ$ 
and 80$^\circ$ mainly near the SR. However, when an EP is not initially located inside an orbital resonance  $\overline{\theta}_{\scriptscriptstyle \text{i}}$ is in average above 50$^\circ$.\\
\subsection{Water transport: merging vs real collisions modelling}
In this section, we compare results of the water transport to EP orbiting inside the HZ, considering two collision models: 
\begin{itemize}
\item [i)] a merging approach in which we consider that the whole water content of the asteroids is delivered to an EP without assuming any water loss processes such as atmospheric drag or ice sublimation. The results using this approach give $\mathcal{W}_{\scriptscriptstyle \text{EP}}$.
\item [ii)] a real collision model using $\overline{v}_{\scriptscriptstyle \text{i}}$ as a key parameter in order to derive the water loss $\omega_{\scriptscriptstyle \text{c}}$ on the asteroid's surface. $\mathcal{\widetilde{W}}_{\scriptscriptstyle \text{EP}}$ contain the results for the water transport from this model.
\end{itemize}
For case ii), we performed detailed simulations of water-rich Ceres-sized asteroids with dry 
targets in the mass range $M_{\scriptscriptstyle \mathrm{EP}}\/$ equal to 1\,$M_{\scriptscriptstyle \mathrm{MOON}}\/$, 
1\,$M_{\scriptscriptstyle \mathrm{MARS}}\/$ and 1\,$M_{\scriptscriptstyle \mathrm{EARTH}}\/$. We assume the Ceres-like 
impactor with $wmf = 15\%$ water content in a mantle around a rocky core. 
The simulations are performed with our parallel (GPU) 3D smooth particle hydrodynamics (SPH) code.
We simulated impacts on targets with different sizes at an impact angle of $30^\circ\/$ 
with initial collision velocities (taken from Fig. \ref{F:impact}) $\overline{v}_{\scriptscriptstyle \text{i}}$ = 2; 5 and 30\,$v_{\scriptscriptstyle \text{e}}$ 
for the Moon,  $\overline{v}_{\scriptscriptstyle \text{i}}$ = 1; 5 and 20\,$v_{\scriptscriptstyle \text{e}}$ for Mars and 
$\overline{v}_{\scriptscriptstyle \text{i}}$ = 1; 3 and 5\,$v_{\scriptscriptstyle 
	\text{e}}$ for the Earth. Except for this latter 
case ($\overline{v}_{\scriptscriptstyle \text{i}}=5\,v_{\scriptscriptstyle \text{e} }$) 
for which about 1 million SPH particles are used,  most of the scenarios are 
resolved in about 500,000 SPH 
particles. All objects were relaxed self-consistently as described in \cite[in prep.]{burmai17}.
\begin{figure}[!h]
	{\includegraphics[width=0.75\columnwidth]{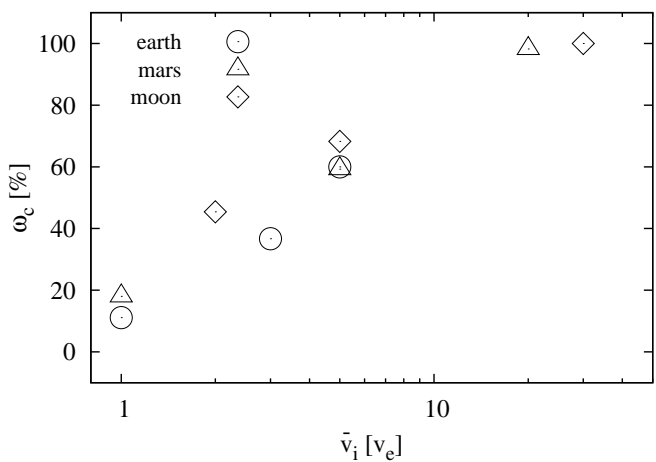}}
	\caption{Water loss $\omega_{\scriptscriptstyle \text{c}}$ as a mass-fraction of the initial total water after impacts on Earth-, Mars- or Moon-sized objets at velocity $\overline{v}_\mathrm{i}\/$ in 
		units of the EP's escape velocity $v_\mathrm{e}\/$.}
	\label{fig:wloss}
\end{figure}
Figure~\ref{fig:wloss} summarizes the water loss $\omega_{\scriptscriptstyle \text{c}}$ in the collision scenarios. All but the most extreme scenario result in a merged main survivor retaining most of the mass. The exception is the $30\,v_\mathrm{e}\/$ impact 
of a Ceres-like body onto the Moon which leads to mutual destruction of the bodies into a debris cloud and hence a loss 
of all volatile constituents. The other very fast impact (Mars at $20\,v_\mathrm{e}\/$) results in a merged survivor 
that retains just under 2\,wt-\% of the available water. For the lower collision velocities we observe water loss rates 
between 11\,wt-\% and 68\,wt-\%. If plotted versus the impact velocity in terms of the mutual escape velocity as in 
Fig.~\ref{fig:wloss}, there is a strong correlation with the impact velocity but only weak dependence on the absolute 
mass.

\begin{figure}[!h]
\centering{
	{\includegraphics[width=0.5\linewidth, angle=-90]{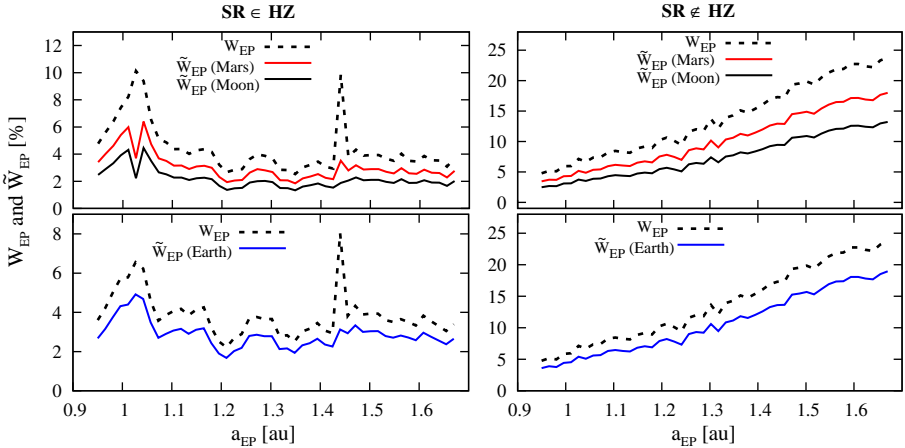}}}
	\caption{Fraction of incoming water with (solid lines) and without (dotted lines) our SPH collision model. \textit{Left} and \textit{right} panels are for SR inside the HZ and inside the asteroid belt, respectively.}
	\label{F:wloss_collision}
\end{figure}

Using a linear extrapolation of $\omega_{\scriptscriptstyle \text {c}}$ between the minimum and maximum impact velocities derived for each EP, we are able to provide better estimates for the water transport. In Fig. \ref{F:wloss_collision} (see also Figs. 2 and 3 in \cite{bancelin17} for more configurations), we compare the fraction of water reaching the EP's surface with and without taking into account our study of SPH collisions represented by the solid and dotted lines respectively (i.e. $\mathcal{\widetilde{W}}_{\scriptscriptstyle \text{EP}}$ and $\mathcal{W}_{\scriptscriptstyle \text{EP}}$) when SR $\in$ HZ or SR $\notin$ HZ (left and right panels, respectively). The two top panels show results for Moon- and Mars-sized objects and the bottom panels, for Earth-sized bodies. We show the comparison for a binary with $q_{\scriptscriptstyle \text{b}}$ = 35 au and a computation of 100 Myr.
One can clearly see that collisions between EPs and asteroids are greatly overestimated if we consider a merging approach. Indeed, we highlight that close to the MMR inside the HZ, the water transport to an EP's surface i.e. $\mathcal{W}_{\scriptscriptstyle \text{EP}}$ can be reduced significantly by almost 50\% for an Earth-, 68\% for a Mars- and 75\% for a Moon-sized object. We have the same statistics near the SR for a Mars- and Moon-sized except for an Earth as $\overline{v}_{\scriptscriptstyle \text{i}} = 3\,v_{\scriptscriptstyle \text{e}}$. In this case, $\mathcal{W}_{\scriptscriptstyle \text{EP}}$ is reduced by $\sim$ 30\%.  Even if no strong perturbation is located in the HZ (right panels), the real collision process shows that the incoming amount of water is reduced to $\mathcal{\widetilde{W}}_{\scriptscriptstyle \text{EP}}\,\sim 50\%\,\mathcal{W}_{\scriptscriptstyle \text{EP}}$ for a Moon-sized EP and to $\sim 20\%\,\mathcal{W}_{\scriptscriptstyle \text{EP}}$ for a Mars- and Earth-sized body (right panels).

A comparison of the left and respective right panels shows that around 1.0 au, $\mathcal{\widetilde{W}}_{\scriptscriptstyle \text{EP}}$ is nearly the same in both cases (SR $\in$ HZ and SR $\notin$ HZ) which indicates the importance of including SPH collisions into dynamical studies to avoid false effects as shown by the dotted lines of left panels in Fig. \ref{F:wloss_collision}:

\begin{itemize}
\item The apparent positive aspect of orbital resonances is that due to eccentric motion of the EP at 1.0 au, the transported water ($\mathcal{W}_{\scriptscriptstyle \text{EP}}$) is boosted and ensures higher values than in case of circular motion of the EP. Moreover, if SR $\notin$ HZ, then even with a higher crossing frequency, the nearly circular motion of EPs close to 1.0 au limits the number of collisions with asteroids.

\item On the other hand, the negative aspect of high eccentric motion near 1.0 au is the high impact velocities which drastically reduces the efficiency of the water transport and leading to significantly  lower values of $\mathcal{\widetilde{W}}_{\scriptscriptstyle \text{EP}}$. It seems that nearly circular motion in the HZ is important to prevent relatively high water loss during collision as $\overline{v}_{\scriptscriptstyle \text{i}}$ are much lower when SR $\notin$ HZ.
 \end{itemize}
 
\section{Conclusions}\label{S:conclusion}
In this work, we investigated the influence of a secondary star on the flux of asteroids into the habitable zone (HZ). We estimated the quantity of water brought by asteroids located beyond the snow line into the HZ of various double star configurations (for which we varied the stellar separation, the eccentricity, and the mass of the secondary). An overlap of perturbations from the secondary and the giant planet in the primordial asteroid belt causes rapid and violent changes in the asteroids' orbits. This leads to asteroids crossing the HZ soon after the gas has dissipated in the system and the gravitational dynamics become dominant. Our results point out that binary systems are more efficient for transporting water into the HZ than  a single star system. Not only an asteroid's flux is 4 -- 6 times higher when a secondary star is present, but also the number of transported oceans into the HZ can be 4 -- 5 times higher, providing other water sources to embryos, in the whole HZ, in the late phase of planetary formation. We highlighted that in tight binaries ($a_{\scriptscriptstyle \text{b}}$ = 50 au), an SR can lie within the inner asteroid belt, overlapping with MMRs, which enables, in a short timescale, an efficient and significant flux of icy asteroids towards the HZ, in which particles orbit in a nearly circular motion. In contrast thereto, in the study of wide binaries ($a_{\scriptscriptstyle \text{b}}$ = 100 au), particles inside the HZ can move on eccentric orbits when the SR lies in the HZ. The outer asteroid belt is only perturbed by MMRs. As a consequence, a longer timescale is needed to produce a significant flux of icy asteroids towards the HZ. This dynamics drastically impacts the dynamical lifetime of particles initially located inside MMRs. It can range from thousands of years to several million years according to the locations of the MMR and whether there is an overlap with the SR. This can favor a fast and significant contribution of MMRs in producing HZc, which are asteroids with orbits crossing the HZ and bearing water therein. In any case, we highlighted that, for the studied binary star systems, the inner disk (region $\mathcal{R}_2$) is the primary source of 
HZc (and therefore for the water in the HZ), by  means of the 2:1 MMR, the 5:3 MMR, and the SR, especially when the latter lies close or beyond the snow line. Finally, we focused on specific binary star -- gas giant configurations where the SR lies either around 1.0 au in the HZ (causing high eccentric motion therein) or inside the asteroid belt beyond the snow-line at 2.7 au (and nearly circular motion in the HZ). We showed that the presence of an SR and an MMR inside the HZ could boost the water transport as the EPs can collide with more asteroids. This is apparently a positive mechanism for the water transport efficiency. However, our study shows clearly that dynamical results overestimate the water transport and need to be corrected by real simulations of collisions which  provides the water loss due to an impact. Indeed, we showed that collisions in the HZ can occur with high impact velocities causing significant water loss for the asteroid (up to 100\%) when colliding at the EP's surface and the amount of water reaching the EP can be reduced by more than 50\%.
%

\acknowledgements{DB, EPL, TIM and AB acknowledge the support of the Austrian Science Foundation (FWF) NFN project: Pathways to Habitability and related sub-projects S11608-N16 "Binary Star Systems and Habitability" and S11603-N16 "Water transport". DB and EPL acknowledge also the Vienna Scientific Cluster (VSC project 70320) 
for computational resources.}

\bibliographystyle{aasjournal}

\bibliography{biblio}

\end{document}